# Atleus : Accelerating Transformers on the Edge Enabled by 3D Heterogeneous Manycore Architectures

Pratyush Dhingra, Graduate Student Member, IEEE, Janardhan Rao Doppa, Senior Member, IEEE, and Partha Pratim Pande, Fellow, IEEE

**Abstract—Transformer architectures have become the standard neural network model for various machine learning applications including natural language processing and computer vision. However, the compute and memory requirements introduced by transformer models make them challenging to adopt for edge applications. Furthermore, fine-tuning pre-trained transformers (e.g., foundation models) is a common task to enhance the model's predictive performance on specific tasks/applications. Existing transformer accelerators are oblivious to complexities introduced by fine-tuning. In this paper, we propose the design of a three-dimensional (3D) heterogeneous architecture referred to as Atleus that incorporates heterogeneous computing resources specifically optimized to accelerate transformer models for the dual purposes of fine-tuning and inference. Specifically, Atleus utilizes non-volatile memory and systolic array for accelerating transformer computational kernels using an integrated 3D platform. Moreover, we design a suitable NoC to achieve high performance and energy efficiency. Finally, Atleus adopts an effective quantization scheme to support model compression. Experimental results demonstrate that Atleus outperforms existing state-of-the-art by up to 56x and 64.5x in terms of performance and energy efficiency respectively.**

**Index Terms—Transformers, Fine-tuning, PIM, Heterogeneity, Pipelining, Accelerators.**

## I. INTRODUCTION

Transformers have emerged as a core technology within the realm of deep learning, buoyed by their success in a huge range of diverse applications, particularly in the domain of natural language processing (NLP) [1]. The recent development of increasingly larger foundation models, driven by the quest for higher accuracy, presents a substantial barrier to the deployment of transformer models at the edge [2]. The training of transformer models with millions of training examples demands substantial computational power and memory requirements [3]. For instance, the training of GPT-2 was conducted using multiple GPUs, utilizing 40 GB of training data and over 8 million web pages [3]. Hence, the training of transformer models from scratch generally requires a server-scale environment. However, foundation transformer models pre-trained using broad and high-coverage data need to be customized to achieve high predictive accuracy for specific applications upon deployment [4]. Some examples include fine-tuning a Large Language Model (LLM) such as GPT for question answering, and fine-tuning LLMs for the finance domain.

Fine-tuning these large-scale language models from scratch is computationally expensive. Traditional full fine-tuning approaches, which require updating all the model parameters via gradient back-propagation, are not feasible and sustainable for resource-constrained edge environments due to a large number of model parameters [5] [6]. Recently, Parameter-Efficient Fine-Tuning (PEFT) has been proposed to fine-tune transformer models efficiently and is shown to achieve comparable accuracy to that of full fine-tuning [7]. PEFT works by updating only a small subset of the model parameters and significantly reduces the computational and memory requirements [4] [8]. However, conventional computing platforms, such as GPUs and TPUs, suffer from suboptimal performance due to the complex dataflow and memory demands imposed by the transformer models even after incorporating PEFT. Quantized pre-trained models along with PEFT have also been explored to alleviate the memory bottlenecks during the fine-tuning process [5]. Quantization helps reduce memory requirements but compromises performance on GPUs. This trade-off arises because data once fetched from memory, is subsequently dequantized for computation to align with the precision requirement of GPUs. Consequently, this scenario motivates us to explore and develop energy-efficient accelerators for fine-tuning transformers.

Processing-in-Memory (PIM) has emerged as a promising approach to accelerate the training/inference of machine learning (ML) workloads [9]. Several non-volatile memory (NVM) technologies, including field-effect transistor memory (FeFET), magnetic random-access memory (MRAM), and resistive random-access memory (ReRAM) have been proposed recently to be used as PIM architectures [10]. These NVM-based PIM architectures can help alleviate the memory demands associated with executing ML workloads including the transformer models. However, transformer models, particularly the attention mechanism need dynamic operand multiplications, which in turn would necessitate multiple write operations to NVM cells. It is well known that NVMs suffer

Received 12 July 2024; revised 6 November 2024; revised 12 December 2024; accepted 13 January 2025. This work was supported in part by the U.S. National Science Foundation under Grant CSR-2308530, and in part by the Army Research Office under Grant ARO-W911NF-24-1-0240. This article was recommended by Associate Editor <>

The authors are with the Department of Electrical Engineering and Computer Science, Washington State University, Pullman, WA 99163 USA (e-mail: pratyush.dhingra@wsu.edu; jana.doppa@wsu.edu; pande@wsu.edu).

from limited write endurance [11]. Further, fine-tuning the transformer model requires additional write operations due to frequent parameter updates during fine-tuning, which can quickly reach the endurance limit for NVM devices [6]. Hence, standalone NVM-based PIM architectures are not suitable for transformer fine-tuning and inference [12]. Therefore, it becomes necessary to consider heterogeneity in the computing substrate to develop a hardware accelerator for transformers to overcome the limitations of standalone PIM or Von Neumann architectures.

- We propose Atleus, a 3D heterogeneous architecture for deploying transformer models on edge. Atleus utilizes NVM and systolic array computing resources to enable low-latency, energy-efficient fine-tuning and inference.
- We design a high-performance heterogeneous 3D NoC in Atleus, which serves as the communication backbone and supports the data exchange among heterogeneous cores.
- Atleus implements a pipelined design across heterogeneous resources that optimizes system utilization and throughput.
- Atleus aligns with the compute and memory demands of transformer models and achieves up to 56× speedup and improves energy efficiency by 64.5× compared to state-of-the-art transformer accelerators.

To the best of our knowledge, this is the first work that proposes a hardware *accelerator for both transformer fine-tuning and inference* to leverage the power of large pre-trained foundation models. The rest of this paper is organized as follows. Section II discusses the prior work, Section III outlines the transformer architecture and fine-tuning method, and Section IV elaborates on the Atleus hardware architecture. Finally, Section V presents the experimental results, and Section VI concludes the paper.

## II. RELATED WORK

In this section, we discuss non-idealities in NVM devices and present relevant existing work on transformer accelerators.

### A. Non-Idealities in NVM Devices

NVM devices are susceptible to non-idealities arising from various factors including process variations, temperature fluctuations, conductance drift, and IR drop, among others [13] [14]. These non-idealities can cause variations in the stored values within NVM cells, potentially leading to erroneous computations and a reduction in the model's predictive accuracy [13] [14]. In this context, several methods have been proposed to mitigate the effect of noise in NVM-based architectures. Among these, retraining in the presence of noise has proven to be an effective strategy for enhancing the robustness of deep neural networks (DNNs). This enables DNNs utilizing non-ideal NVM devices to achieve accuracy similar to their noise-free counterparts for both training and inference. While noise-injection approaches improve robustness to NVM non-idealities, it does not address the limitation of low write endurance in NVM devices. The write endurance of NVM devices typically ranges from $10^6$ to $10^{12}$ writes [11]. The parameter update necessary for finetuning or training would require frequent writes to NVM cells which will lead to device failure. Existing approaches aimed at reducing the number of rewrites during training or fine-tuning incur significant performance overhead or necessitate redundant hardware [15] [16]. In this work, we finetune transformer models while preventing frequent rewrites to NVM crossbars by leveraging heterogeneous resources for model acceleration.

### B. Transformer Accelerators

Currently, GPUs are predominantly used to accelerate transformers [1]. Google's Tensor Processing Units (TPUs) based on systolic arrays is another alternative for accelerating transformers and various other machine learning models [17] [18]. However, both GPUs and TPUs suffer from low utilization with computing cores waiting for data due to limited memory bandwidth [17] [19]. Additionally, the cost of data transfer is orders of magnitude higher than computation, leading to suboptimal performance and poor energy efficiency [20]. This has led to research into alternative hardware platforms to accelerate transformer models. These include a diverse range of implementations using FPGA, PIM, and different combinations of NVM-based PIM and SRAM blocks. ReTransformer proposes a ReRAM-based PIM architecture to accelerate transformer inference [21]. AccelTran also proposes a 3D ReRAM-based PIM accelerator proposed for edge deployment [22]. Similarly, another work introduces a hardware-software co-design framework for transformer acceleration on ReRAM-based architecture [23]. As mentioned earlier, PIM as a standalone solution is not suitable for both transformer fine-tuning and inferencing given the write endurance problem of NVM devices. FTrans is an FPGA implementation for transformer inference but has high energy overhead [24]. TransPIM is a DRAM-based PIM accelerator with compute units integrated within High Bandwidth Memory (HBM) banks to accelerate transformer inference [25]. TransPIM uses the host device connected via an interposer for non-matrix computations (normalization, activation functions, etc.) that cannot be performed within the DRAM itself. This off-loading of computations adds a significant latency overhead and leads to low overall system utilization. Addressing this non-matrix computation latency, HAIMA adopts a hybrid strategy, by incorporating SRAM units for dynamic computations in

Table I. Comparison b/w existing transformer accelerators and Atleus

| Ref. | Heterogeneous | Inference/ Fine-Tuning | Mixed Precision | NoC Design |
|------|---------------|------------------------|-----------------|------------|
| GPU | N | Y/Y | Y | Y |
| TPU | N | Y/Y | Y | Y |
| [21] [22] | N | Y/N | N | N |
| [24] | N | Y/N | N | N |
| [25] | N | Y/N | N | N |
| [26] | Y | Y/N | N | N |
| [27] | Y | Y/N | N | N |
| *Atleus* | **Y** | **Y/Y** | **Y** | **Y** |

ª. "Y" represents suitability, while "N" denotes unsuitability for the technique.

self-attention and DRAM for multiplications involving large weight matrices [26]. Likewise, H3D-Transformer proposes a hybrid architecture consisting of FeFET, SRAM, and TPU cores stacked vertically via TSVs in a 16-tier system [27]. SwiftTron is an accelerator with custom multiply-and-accumulate (MAC) PEs designed to execute transformer computational kernels [28]. However, it is important to note that existing works in academia like HAIMA, TransPIM, SwiftTron, and H3D-Transformer demonstrate performance gain only for inference and do not consider fine-tuning. The transformer fine-tuning is a more challenging task characterized by complex data dependencies. Further, existing works do not consider communication traffic while integrating manycore architecture. The NoC framework integrating the computing cores necessitates optimization to facilitate efficient data exchange and support compute parallelism. In this paper, we fill this gap in the state-of-the-art by proposing a heterogeneous multi-core 3D architecture aimed at enabling transformer models to run efficiently for both fine-tuning and inference on the edge. Table I summarizes and compares a few of the existing notable transformer accelerators from industry and academia with respect to our proposed Atleus architecture.

## III. BACKGROUND ON TRANSFORMERS

In this section, we provide relevant background on transformer architectures and elaborate on relevant fine-tuning methods proposed for transformers.

### A. Computation Kernel

The transformer architecture is composed of multiple sequential layers of encoder/decoder blocks. Each of these blocks consists of two major functional modules: multi-head attention (MHA) and feed-forward (FF) network [1]. The MHA module employs multiple attention heads with each head computing attention scores [1]. Conversely, the FF module is a fully connected network consisting of two linear layers designed to process the data sequentially [1]. Table II outlines the transformer computational kernels along with the associated operations. Here, $d_{model}$ represents the model dimensionality of the transformer architecture and $n$ represents the input sequence length. We use the same terminology conventionally used to describe transformer architecture. For instance, $Q$, $K$, $V$, and $S$ represent the query, key, value, and attention score vectors. The MHA computation involves multiplication with four trainable weight matrices ($W^Q, W^K, W^V$, and $W^O$) each with dimension $d_{model} \times d_{model}$. The other MHA computations include multiplications with dynamic operands and non-linear softmax and normalization operations. The FF network is bigger in comparison to MHA with $W^{F1}$ and $W^{F2}$ trainable weight matrices having dimensionality ($d_{model} \times d_{ff}$), where $d_{ff} = 4d_{model}$. Each computation module is followed by layer-normalization to ensure stability.

### B. Fine-Tuning

Transformer models require fine-tuning to enhance their predictive accuracy for specific tasks such as text classification, question-answering, translation, etc. There exist several PEFT techniques to fine-tune transformers efficiently. These methods encompass a diverse set of approaches; including adding new trainable layers, optimizing input layer activations, making a subset of parameters trainable, updating input layer embeddings, introduction of biases, and prompt tuning [29] [30] [31]. Low-rank Adaptation (LoRA) is a state-of-the-art fine-tuning technique that works by introducing low-rank matrices to modulate the weights of the pre-trained model [6]. Only the low-rank matrices are updated during the fine-tuning process, keeping the original model's parameters unchanged. Eq. 1 shows the LoRA computation kernel.

$$Y = WX = (W_O + AB)X = W_O X + ABX \qquad (1)$$

Here, $Y$ is the output with input being $X$, and $W$ is the weights after fine-tuning. The fine-tuned weights are the sum of the pre-trained weights ($W_O \in R^{d \times d}$) which were frozen, and the low-rank matrices product ($AB$). $A \in R^{d \times r}$, $B \in R^{r \times d}$ are the two trainable low-rank matrices where "$r$" is the rank and "$d$" represents the pre-trained weight matrix dimension. These low-rank matrices are much smaller in size compared to the original weight matrix with $r \ll d$ [6]. Mathematically, this changes the computational complexity of the trainable parameters from $O(d^2)$ to $O(d.r)$, thereby significantly reducing the number of matrix multiplication operations. For instance, existing works have shown that the value of $r$ in the range of 8 to 32 is sufficient for fine-tuning a model with $d_{model} = 2048$ [6]. Fig.1 shows the transformer encoder block with low-rank adapters added for the MHA module. Note that LoRA can be applied to any of trainable weight matrices within the transformer model.

Table II. Transformer Computational Kernels

|  | **Multi-Head Attention (MHA)** | **Operations** |
|---|---|---|
| **MHA-1** | $Q, K, V = X.W^Q, X.W^K, X.W^V$ | $3 \times O(d_{model}^2 .n)$ |
| **MHA-2** | $S = Q.K^T$ | $O(d_{model}.n^2)$ |
| **MHA-3** | $P = Softmax(S).V$ | $O(d_{model}.n)$ |
| **MHA-4** | $H_m = P.W^O$ | $O(d_{model}^2 .n)$ |
| **L-1** | $M = LayerNorm(X + H_m)$ | $O(d_{model}.n)$ |
|  | **Feed-Forward (FF)** |  |
| **FF-1** | $X^1 = GeLU(M.W^{F1})$ | $4 \times O(d_{model}^2 .n)$ |
| **FF-2** | $X^2 = GeLU(X^1.W^{F2})$ | $4 \times O(d_{model}^2 .n)$ |
| **L-2** | $X_{new} = LayerNorm(M + X^2)$ | $O(d_{model}.n)$ |

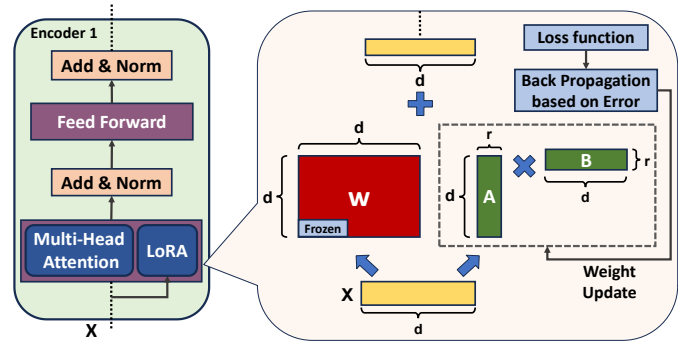

Fig. 1: High-level illustration of the transformer encoder with low-rank adapters added to the MHA module for fine-tuning.

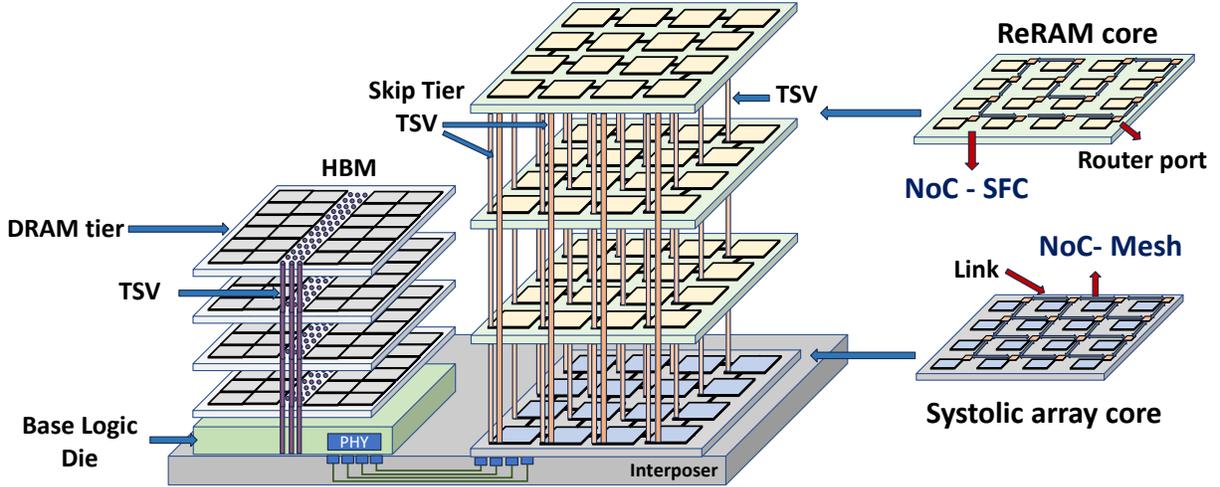

**Fig. 2:** Atleus architecture consisting of 3 ReRAM tiers and 1 systolic array tier connected vertically through TSV-based interconnect. The DRAM is connected through the interposer via 2.5D integration. This figure is for illustration purposes only.

## IV. ATLEUS ARCHITECTURE

In this section, we introduce the Atleus heterogeneous manycore architecture and present our overall design strategy. First, we discuss the computational kernel to core mapping. Subsequently, we discuss the design of the NoC architecture considering the communication traffic pattern while executing the transformer model. Next, we describe the systolic array design based on the computational kernel mapped during transformer fine-tuning and inference. Afterward, we elaborate on the quantization of the transformer model within Atleus and discuss the overall manycore architecture.

### A. Compute Mapping

The transformer model incorporates two distinct categories of matrix multiplication (*MM*) operations: 1) involving trainable weights (MHA-1, MHA-4, and FF-1, FF-2); and 2) with dynamic operands (MH-2, MHA-3). *MM* with static weights can be performed effectively on NVM devices. NVM crossbars perform $N^2$ multiplications in $O(1)$ time making it extremely energy efficient and achieving high performance [20]. Therefore, we propose to utilize ReRAM-based PIM for these computations since it is the most compact/dense NVM, making it well-suited for edge computing. We refer to the *MM* operations performed by ReRAM as $MM_{ReRAM}$. The proposed manycore architecture and associated design strategy is also applicable to other NVM devices. Further, we propose to employ systolic arrays for the *MM* tasks involving dynamic operands as well as for additional computations required during the fine-tuning phase, such as LoRA. Systolic arrays consist of processing elements (PEs) connected in a grid where data is sent sequentially through the edges which then propagates among the PEs [32]. The reuse of data within the grid helps in reducing memory bandwidth and makes them suitable for *MM* operations [32]. We denote the *MM* operations performed by the systolic array as $MM_{systolic}$.

We determine the value for the parameters $MM_{ReRAM}$ and $MM_{systolic}$ defined above for the two computing resources based on the computational complexity presented in Table II. ReRAM cores are exclusively utilized for the FF computation as well as

MHA computation involving trainable weights. Hence, the parameter $MM_{ReRAM}$ will be as given below:

$$MM_{ReRAM} = 12 \times O(d_{model}{}^2 . n) \qquad (2)$$

Here, $MM_{ReRAM}$ is the summation of MHA-1, MHA-4, FF-1 and FF-2 operations. Conversely, systolic array cores are utilized for dynamic computation in MHA, non-linear operations such as softmax and layer normalization, and LoRA computation if considering the fine-tuning phase.

$$MM_{systolic} = O(d_{model} . n^2) + 2k \times O(d_{model} . r . n) + \\ + 3 \times O(d_{model} . n) \qquad (3)$$

The $MM_{systolic}$ includes the operations MHA-2, LoRA, and non-linear computations MHA-3, L1, and L2. Here, *k* denotes the number of matrices where LoRA is applied. The term $2k \times O(d_{model} . r . n)$ in Eq. 3 represents the LoRA computational complexity, where the factor $2k$ accounts for both the forward and backward propagation phases. Note that the computational complexity remains the same for both the propagation phases. Additionally, the term $3 \times O(d_{model} . n)$ in Eq. 3 represents the computational complexity of all three non-linear computations (MHA-3, L1, and L2). We later show in our experimental evaluation that ReRAM is utilized for more than 90% of the *MM* for both fine-tuning and inference.

Overall, the advantages of using heterogeneous computing resources for transformers are two-fold. First, we perform the weight-stationary computation that represents most of the *MM* operations in an energy-efficient manner by utilizing ReRAM. Second, we mitigate the endurance issue of ReRAMs by employing systolic arrays whenever dynamic operands are involved. Fig. 2 shows the Atleus architecture consisting of vertically integrated tiers, each with either ReRAM or systolic array cores. The tiers are connected vertically with Through Silicon Via (TSV)-based links using hybrid bonding [33] [34] [35]. The off-chip access to DRAM is enabled through an interposer via 2.5D integration [36]. The DRAM used is High Bandwidth Memory (HBM), which is a multi-layered DRAM technology specifically designed to offer high bandwidth for data-intensive applications such as the transformer acceleration considered in this paper. An industry standard interface protocol (DFI) is considered between the DRAM and Atleus [37]. It generates all handshake signals with precise timing requirements necessary for interfacing.

Generally, NVM-based PIM architectures adopt a pipelined design during batch-wise neural network (NN) training to optimize throughput [38]. Inspired by this, we incorporate a pipelined design for transformer fine-tuning as well. However, it is critical to ensure that the pipeline does not break across heterogeneous computing resources. Unlike GPUs, both ReRAM and systolic array execute instructions in-order with deterministic latencies. Hence, we employ a deterministic model to evaluate the execution time of each computation kernel and ensure a pipelined implementation. Like other NNs, we create a pipeline across encoders/decoders (layers) of transformers. However, unlike other NNs, each layer within the transformer involves several computational kernels. This provides the possibility of employing a more granular pipeline within each layer to maximize system utilization. Hence, we propose to implement an intra-layer pipeline in Atleus to further increase resource utilization and efficiency.

The compute mapping distributes the intra-encoder/decoder computation across the heterogeneous resources of the Atleus manycore architecture. The $W^Q, W^K, W^V$ matrices are stored on ReRAM cores, which generate the $Q, K$ and $V$ vectors. These vectors are sent to the systolic array to compute the attention score. We adopt the method of fused score and softmax calculations [39]. The softmax values are computed row-wise and appropriately scaled to get the actual softmax for the attention head. This ensures that the attention values are computed without the need to write intermediate matrices back to DRAM, preventing frequent DRAM data access. The fine-tuning computations (LoRA) are also performed on the systolic array. Following MHA, the activations are transferred to the ReRAM for FF network processing. An effective pipelined implementation requires layers of the transformer models to be executed simultaneously. ReRAMs perform in situ MAC operations and necessitate the MHA and FF pre-trained weights on separate cores for pipelining [38]. Conversely, there is no such requirement for systolic arrays. It can load parameters to execute computational kernels within a pipeline stage. Since ReRAM is utilized for most of the matrix operations, and systolic array is employed to perform only dynamic computations; a pipelined design requires more ReRAM cores compared to systolic array cores.

### B. NoC Design Structure

Recall that Atleus is a 3D manycore architecture. It is necessary to ensure that cores communicating with each other are mapped nearby to facilitate low-latency communication. The physical distance between cores in the vertical direction is less when compared to the horizontal direction [40]. Consequently, we propose to map the intra-layer computing cores in the vertical direction to minimize the communication latency. However, as discussed earlier, the number of ReRAM cores outnumber the systolic array cores. Note that both the systolic array and ReRAM cores cannot be integrated on the same tier due to technology limitations. Hence, more tiers are allocated to ReRAMs than systolic arrays. This vertical mapping would necessitate bypass links between ReRAM and systolic array cores where we may have to skip one ReRAM tier to communicate directly with the systolic array. Therefore, we need long-range shortcuts between the ReRAM tier and the systolic array tier to prevent multi-hop communication. Atleus employs TSV-based long-range links to enable energy-efficient skip-tier communication. However, the aspect ratio of a TSV, defined as the ratio of its depth to its diameter, directly correlates to stress-related reliability [41]. Consequently, longer skip-tier TSVs necessitate a corresponding increase in TSV diameter. We analyze the area and performance trade-offs associated with the incorporation of skip links in the experimental results section.

Next, we determine the core and link placement across transformer layers. The heterogeneous architecture in Atleus consists of ReRAM and systolic array tiers. These two technologies necessitate distinct interconnect strategies. The ReRAM crossbars store the pre-trained weights on-chip, which are spatially distributed across ReRAM cores. The resulting activations flow from the $i^{th}$ layer to the $(i + 1)^{th}$ layer in a unidirectional manner across ReRAM cores. If two consecutive transformer layers are mapped far apart, it will lead to long-range multi-hop communication. This, in turn, will degrade the performance and energy efficiency of the NoC. Therefore, it is essential to ensure continuity in the ReRAM core placement to the extent possible to reduce communication overhead. Existing work has shown that employing a space-filling curve (SFC) for 2.5D chiplet systems ensures contiguity between any two consecutive neural layers and minimizes the communication overhead [42]. Thus, we utilize and adapt the concept of SFC to connect the cores in each ReRAM tier. Fig. 2 highlights the interconnection on the ReRAM tier where cores are connected along the path formed by the SFC. The incorporation of an SFC ensures single-hop connectivity for data exchange, unlike conventional topologies such as mesh or torus. Unlike in ReRAM, where only activations flow, both input data and activations flow orthogonally across cores in the systolic array tier. Additionally, the systolic array necessitates data exchange from DRAM for computations. Hence, we employ a conventional mesh-based NoC to effectively manage the traffic on the systolic array. A standard NoC flow control mechanism (FIFO-based) is utilized for synchronization across heterogeneous cores. Fig. 2 shows the NoC design in Atleus.

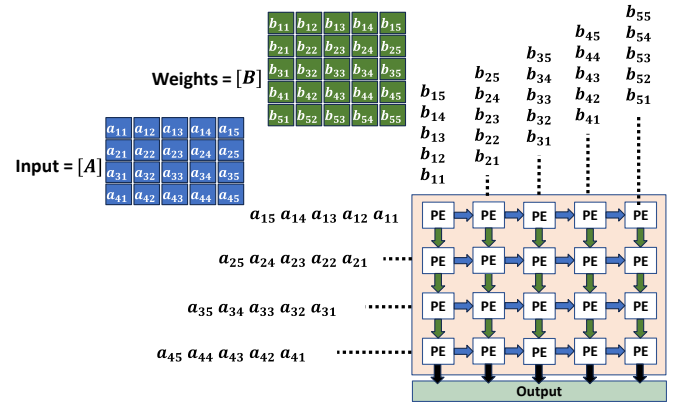

**Fig. 3:** Illustration of systolic array-based $MM$ between input matrix [A] and weight matrix [B] using OS dataflow. Each PE retains the partial sum, while the input is transferred to the rightward PE, and the weight is moved to the downward PE each cycle.

### C. Systolic Array Design

*MM* within NNs typically involves the multiplication of the input vector with the weights to generate the output matrix. There exist three strategies for mapping a computational kernel onto systolic arrays: namely Output Stationary (OS), Input Stationary (IS), and Weight Stationary (WS) [32]. The "stationary" term determines which vector among the input, weight, or output remains fixed within the PE during the computation. Fig. 3 shows the OS data flow in the systolic array as an example. The selection of a configuration plays an important role as it influences the total number of compute cycles. The WS/IS dataflow involves preloading the stationary matrix onto the PEs prior to the computational tasks, making them suitable for scenarios where one of the matrices is reused repeatedly. Conversely, the OS dataflow is optimal when the *MM* involves dynamic operands as it loads both inputs simultaneously, thereby reducing pre-filling cycles [43]. Given that we map dynamic operand multiplications on the systolic array during transformer fine-tuning and inference, we propose to implement the OS dataflow in Atleus.

Next, we need to determine the suitable dimensions for the grid of PEs in the systolic array. The systolic array dimensions, characterized by its length and width, directly influence the array's capability to effectively execute a given workload. We consider different array dimensions and determine the configuration that achieves the highest average utilization on the various computational kernels mapped to the array during fine-tuning. Later, we show in our experimental evaluation that a rectangular systolic array achieves the best cumulative performance on the computational kernels allocated to the systolic array. This outcome diverges from conventional implementations of systolic arrays, which adopt a square configuration, e.g., Google TPUs [18]. The primary reason for the enhanced performance of the rectangular configuration can be attributed to the nature of the computational kernels mapped onto the array. Specifically, LoRA computations exhibit a distinctly rectangular matrix with $r \ll d_{model}$, as mentioned in Section III. This renders square configurations suboptimal with low utilization when mapped. Consequently, we adopt a rectangular dimension for the systolic array within Atleus aligning with the computational kernels during fine-tuning.

### D. Crossbar-Wise Quantization

Quantization is a commonly used model compression technique that works by transforming the input to a lower-bit representation through a discretization process. Recently, QLoRA (Quantized Low-Rank Adaptation) has been proposed for transformer fine-tuning. QLoRA extends the principle of LoRA by quantizing the pre-trained model [5]. It employs two precision types, a low-precision storage format for the weights and a high-precision format for computing. Eq. 4 shows the QLoRA computation kernel.

$$Y = W_O^{Quantized} X + ABX \qquad (4)$$

Here, the pre-trained model weights are quantized, and the low-rank matrices are kept at higher precision to preserve accuracy. Block-wise $k$-bit quantization technique is utilized to achieve higher compression of the pre-trained model [44]. The approach works by segmenting the input into distinct blocks and quantizing each block independently with a scale value. The quantization scale is later utilized to dequantize the value during computation. The primary reason for adopting quantization on GPUs is to reduce off-chip data access and reduce the memory requirement during fine-tuning. The quantized blocks once fetched from memory need to be dequantized since GPUs do not support arbitrary bit precisions. This tradeoff for lower memory requirement results in repeated cycles of quantization-dequantization of data, thereby adding computational overhead. Conversely, there is no such limitation for NVM-based architectures. NVM crossbars support random bit precision with no dequantization necessary for computation. Additionally, quantization on NVM crossbars helps reduce the number of cells needed to represent each weight. For instance, the quantization of a 32-bit input into 8-bit requires 25% of the original cells needed for storage. This reduction directly results in either reduced resource requirements or faster-pipelined execution with weight duplication. However, NVM-based architectures require uniform bit representation within each crossbar; vectors stored across multiple cells must maintain consistent bit representation to ensure correct MAC operation.

We extend the principle of block-wise quantization from QLoRA and propose crossbar-wise quantization in Atleus. The method works by quantizing the pre-trained weights on each ReRAM crossbar independently. However, there is a challenge when the weights of the NN are spread across multiple crossbars. The activations need to be aggregated to obtain the final output for matrix-vector multiplication (MVM). The direct summation of activations, which may be quantized with different quantization scales, will result in incorrect aggregation. Consequently, activations from each crossbar need to be dequantized before aggregation. Note that this need for dequantization is different from GPUs, where dequantization precedes computation. *In contrast, our proposed method involves dequantizing the output product post MVM*. This yields in a substantial reduction, by a factor equivalent to the crossbar size (128 for Atleus), in the frequency of dequantization operations compared to GPU architectures. Fig. 4 highlights the difference in the need for dequantization on both platforms. Existing ReRAM-based PIM architectures do not support dequantization at the crossbar level. Therefore, we modify the existing ReRAM architecture design to support dequantization. The dequantization process, which involves multiplication with a constant scalar, can be efficiently achieved on hardware through shift-and-add or shift-and-subtract operations. We facilitate dequantization by integrating additional shift-and-add (S&A) units into the previously proposed ReRAM tile architecture [20]. We do not utilize existing units as additional S&A operations introduce timing overhead. This, in turn, will make existing (S&A) units the bottleneck in the pipelined implementation. We also incorporate additional registers dedicated to storing the quantization constants to dequantize the value. Conventionally, the current through the crossbar bit lines is converted to digital values through an ADC. Subsequently, the output is obtained by performing bitwise S&A operations and stored in the output register. Our proposed modification introduces an extra computation cycle in this process. Following the initial conversion, S&A operations are executed to dequantize the value utilizing the quantization constant. Fig. 5 shows the MVM dataflow on ReRAM crossbars with the additional pipeline stage used for dequantization. The inclusion of dequantization capabilities directly within the ReRAM tile enables the execution of models with crossbar-wise quantization. The architectural modifications introduce negligible area and power overheads of 2.15% and 1.5% respectively.

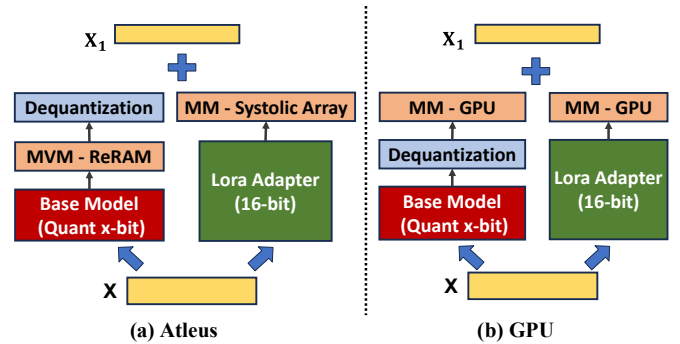

**Fig. 4:** Conceptual illustration showing dataflow for fine-tuning on (a) Atleus and (b) GPU, using a quantized model. The MVM on ReRAM uses the quantized model and dequantization occurs post computation, unlike GPUs where dequantization precedes computation.

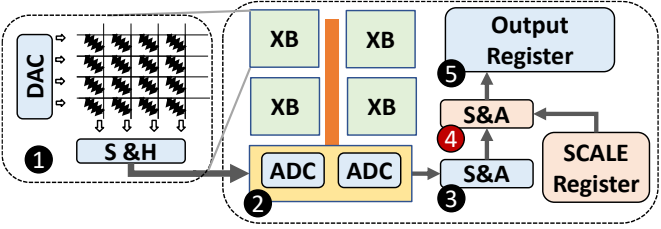

**Fig. 5: MVM dataflow in ReRAM tile architecture with the integration of additional S&A for dequantization.**

## V. EXPERIMENTAL RESULTS

In this section, we present an experimental evaluation of the Atleus architecture. First, we describe the experimental setup used for the performance evaluation. Subsequently, we discuss the systolic array configuration in Atleus. Next, we analyze the computational breakdown among ReRAM and systolic array. Afterward, we discuss the NoC configuration in Atleus and detail the cost analysis of Atleus. Next, we consider non-idealities in ReRAM. Thereafter, we present a detailed performance and energy analysis with respect to existing transformer accelerators. Next, we evaluate the effectiveness of quantization on Atleus and compare it with the baselines. Finally, we assess Atleus on inference tasks.

### A. Experimental Setup

We consider four transformer models in our experimental evaluation: Roberta-Base, BERT-Large, GPT-2 (Medium), and BLOOM-560m [45]. The models represent a broad range of model parameters from 110 million to 560 million. These transformer models are fine-tuned on three real-world datasets [46]. Table III provides the nature of tasks, the dataset, and the number of training examples for fine-tuning. We employ the popular LoRA technique on the $W^Q$ and $W^V$ pre-trained weight matrices for transformer fine-tuning with rank $(r) = 32$, following existing work [6]. Note that LoRA can be applied to other pre-trained weights and $r$ is a hyperparameter. The systolic array was synthesized using the *Synopsys Design Compiler*. Additionally, the power is calculated via *Synopsys Spyglass*. For the ReRAM tiers, each core consists of 16 tiles. We modify the existing ReRAM tile configuration to incorporate crossbar-level dequantization [20]. The area and power specifications of the ReRAM and systolic array cores used are shown in Table IV. We utilize industry standard HBM2 as the DRAM connected through the interposer for off-chip data access. Table IV presents the HBM2 parameters.

The system-level evaluation considers both computation and communication for reporting latency and energy metrics. The latency of each core is computed using cycle-accurate simulators. *SCALE-Sim* is utilized for the systolic array cores and a modified *NeuroSim* is employed to obtain the latency of all on-chip buffers, and peripheral circuits in ReRAM cores [47] [32]. Afterward, we use the cycle-accurate *BookSim2* simulator to implement the NoC to connect the cores and determine NoC latency and power [48]. The inputs to the *BookSim2* are the connectivity between routers, the inter-core traffic, and the technology file. We utilize the TSV link design parameters from existing literature [33]. Table IV highlights the physical parameters for the TSV interconnect. In our performance evaluation, we assume that the pre-trained model parameters are mapped to ReRAM crossbars prior to inferencing or fine-tuning consistent with prior work [38]. For systolic arrays, we take into account the timing overhead associated with loading weights from the DRAM.

As discussed in Section IV, a pipelined approach is utilized to execute each layer in Atleus. We balance the delay across the multiple stages in the intra-layer pipeline to ensure high throughput. The MHA computation with pre-trained weights is executed in a single stage on the ReRAM cores. Afterward, the systolic array is utilized in the next stage to compute dynamic vector computation. Next, the FF network is implemented on ReRAM cores. The FF network has twice as many pre-trained parameters as compared to MHA (see Table II). Hence, the FF network is divided into two stages given that the throughput of the pipelined system is constrained by its slowest component. Overall, the intra-layer pipeline consists of four stages with ReRAM utilized in three of these stages and systolic array utilized in the remaining stage. We allocate resources accordingly in the ratio of 3:1 among ReRAM and systolic array cores. The 3D system under consideration is structured with four planar tiers, each with size 10 mm x 10mm. We utilize 3 ReRAM tiers and 1 systolic array tier, each with 16 cores placed in a 4 x 4 grid as shown in Fig. 2. TSV interconnect is utilized to connect planar tiers with 48 vertical links. Additionally, the bottom tier (systolic array) and the top tier (ReRAM) are connected point-to-point through 16 TSV links to support long-range communication. We maintain a consistent aspect ratio for all TSV links including skip links. Hence, for skip links, we need to increase TSV diameter. We consider this area overhead in our analysis. Fig. 2 shows the NoC in Atleus incorporating TSV links for adjacent-tier and skip-tier communication. Note that this is an example system size for the performance evaluation. We compare the performance of the Atleus with the recently proposed state-of-the-art transformer accelerator, HAIMA. To ensure consistency, we maintain the same DRAM data access and energy parameters across our evaluation. As a second baseline, we employ a 3D-systolic array (3D-TPU) architecture based on the Google TPU-v4 [18]. The 3D-TPU architecture is a four-tier system featuring 128x128 systolic array connected via a 3D-torus interconnect [18]. Each tier contains 4 cores placed in a 2 x 2 configuration, maintaining comparable tier area to that of Atleus. Additionally, we utilize the same clock frequency and the amount of on-chip SRAM to align with Atleus's systolic array configuration for fair evaluation. We also consider a Nvidia Tesla V100 GPU for comparative performance evaluation. For the GPU implementation, we utilize PEFT, LoRA, and QLoRA library from Hugging-Face [49]. The GPU is ensured not to be in power-saving mode by running dummy matrix multiplication prior to fine-tuning or inferencing transformer workloads. We do not compare with homogenous ReRAM-based PIM architecture, ReTransformer or Acceltran in this comparative performance evaluation due to the endurance concern [21] [22]. To illustrate, consider the BERT-Large model processing a single input sequence of length $n = 1024$, where each attention head is mapped to a unique ReRAM core. We need ~$5 * 10^4$ rewrite operations to ReRAM cells. Notably, the number of necessary rewrites will increase while fine-tuning due to training of LoRA parameters on ReRAM cores.

Table III. Fine-Tuning tasks and dataset

| Task | Dataset | Training Examples |
|------|---------|-------------------|
| Question-answering | Squad | 130319 |
| Multiple-choice | Swag | 73546 |
| Language-Modeling | Wikitext | 36718 |

Table IV. Atleus Heterogeneous Architecture Specifications

| ReRAM Tier: 16-cores, 16-tiles per core | |
|---|---|
| ReRAM Tile | 96-ADCs (8-bits), 12×128×8 DACs (1-bit), 96 crossbars, 128×128 crossbar size, 2-bit/cell resolution, 0.345 W, 0.37 mm², 12 Scale Register (192B), 12×4 (S&A) units, Tech node – 32 nm |
| **Systolic Array Tier: 16-cores** | |
| Systolic Core | 128×32 array of PEs, Total SRAM – 1 Mb, 800 MHz clock, 2.55 mm², 2.13 W, Tech node – 10 nm |
| **NoC Design Parameters** | |
| TSV Parameters | Diameter – 5 μm, Via Height – 15 μm, Capacitance – 37 fF, Resistance- 20 mΩ [33] |
| DRAM Parameters | 2 GB per tier, 256 GBytes/sec max bandwidth, 1024-bit memory interface, 600 MHz clock [25] |

### B. Systolic Array Design

First, we determine the systolic array size based on the computational kernels mapped during transformer model fine-tuning. Unlike ReRAM, systolic array cores execute multiple computations in a single pipeline stage, which include attention score, softmax, normalization, and LoRA computations. It is necessary to ensure that the systolic array does not become the bottleneck given the slowest pipeline stage determines the achievable performance. The grid size within each systolic array core directly correlates to the overall compute capability. A larger grid of PEs facilitates faster execution but also increases power consumption. Consequently, it is crucial to identify the minimum number of PEs necessary to execute the computations so that the pipeline is balanced. Further, it is also important to determine the dimensions of the grid that maximize the overall utilization on the computational kernels mapped. A higher utilization ensures an energy-efficient implementation. Fig. 6 illustrates the normalized delay associated with varying numbers of PEs arranged in different grid dimensions for two transformer models as an example. Here, the delay for each computational kernel is normalized to the slowest ReRAM tier stage delay in the pipeline. The input sequence length used is the maximum allowed in each model to consider the worst-case delay. We observe that smaller arrays with 1024 or 2048 PEs are not sufficient to execute the computational kernels in a single pipeline stage and 4096 PEs are necessary to meet the delay constraints. A configuration with 4096 PEs and dimensions of 128x32 is experimentally found to be the most efficient, balancing compute resources and delay for the models considered in this work. This demonstrates the effectiveness of our systolic array configuration for the application under consideration.

### C. Computational Breakdown: ReRAM vs. Systolic Array

Now, we consider the comparison between the ReRAM and systolic array cores in terms of compute load and energy consumption during the fine-tuning process. Figs. 7(a) and 7(b) illustrate the computational and energy breakdown, respectively with GPT-2 (Medium) as an example. Note that the findings and pattern of results are similar for other transformer models used in our evaluation. We utilize the maximum allowable sequence length for GPT-2 (Medium) which is 1024 for this analysis. From Fig. 7(a), we observe that Atleus utilizes NVM cores for 91.9% of the computations. However, as shown in Fig. 7(b), the energy contribution of the ReRAM is less than that of systolic arrays. This highlights the significant energy savings enabled by the ReRAM cores, given that they perform

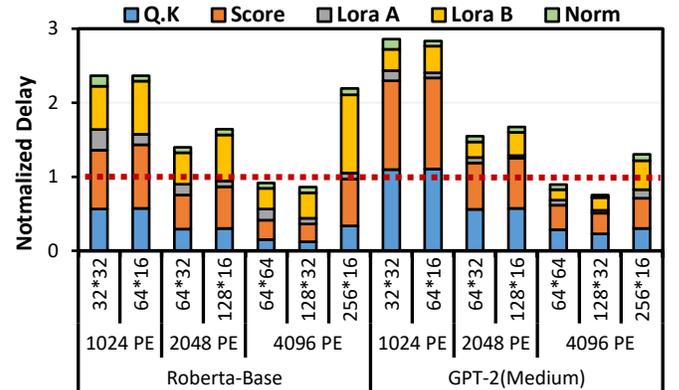

**Figure 6: Normalized delay for each computational kernel mapped onto the systolic array with respect to ReRAM tier pipeline stage delay. Here, LoRA A and LoRA B represent the low-rank adapter layers introduced to the $W^Q$ and $W^V$ matrices, as shown in Eq. 1.**

approximately 11.34× more operations compared to systolic arrays.

Next, we determine the ratio of ReRAM to systolic array cores required to implement the *MM* computations analytically. We calculate the *MM* for each type of core as defined in Eqs. 2 & 3 in Section IV. The ratio is computed by dividing the *MM* of ReRAM by those performed by the systolic array. The ReRAM to systolic array *MM* can be expressed as follows:

$$\frac{MM_{ReRAM}}{MM_{Systolic}} \propto O\left(\frac{12 \times d_{model}^2 . n}{d_{model}. n^2 + 2k \times d_{model}. r.n + 3 \times d_{model}. n}\right)$$

Here, $d_{model}, n, r$ and $k$ represent the model dimension, input sequence length, rank for the LoRA matrices, and the number of LoRA matrices in each layer, respectively. For the scenario considered here, k = 2. After simplifying the expression, we obtain:

$$\frac{MM_{ReRAM}}{MM_{Systolic}} \propto O\left(\frac{12 \times d_{model}}{n + 4 \times r + t_{const}}\right)$$

The non-linear computations such as softmax and layer-normalization scales linearly with both $d_{model}$ and $n$, thus introducing a fixed computational overhead $t_{const}$ to the overall execution time irrespective of $d_{model}$ or $n$. Given that r << $d_{model}$, the LoRA computation does not impact the ratio as well. Consequently, the ratio of ReRAM and systolic array computations can be approximately expressed as:

$$\frac{MM_{ReRAM}}{MM_{Systolic}} \propto O\left(\frac{12 * d_{model}}{n}\right) \quad (5)$$

For instance, in GPT-2 (Medium), where $d_{model} = 1024$ and maximum allowable sequence length ($n_{max}$) = 1024, ReRAM is utilized for ~12× the computations compared to the systolic array. This observation aligns with the results shown in Fig. 7 (a), which illustrates the breakdown of ReRAM and systolic array computations. Overall, the ReRAM to systolic array ratio varies between 90.08% to 94.7% based on $d_{model}$ and $n_{max}$ of the transformer models considered in our work. This underscores the effectiveness of Atleus which leverages the advantages of ReRAM-based PIM in reducing system energy for more than 90% of the computations, while preventing frequent writes on crossbars during fine-tuning.

### D. NoC Optimization

Next, we evaluate the performance of the NoC incorporated in Atleus. We examine three distinct NoC designs in terms of EDP, area, and cost for this evaluation:

1) **3D-mesh:** A 3D mesh enabled by TSV links. It serves as a reference point for comparison.
2) **3D-mesh with skip:** An extension of the 3D-mesh design where we introduce additional TSV-skip links connecting the top and bottom tier. The additional links serve as shortcuts to the mesh design to support intra-layer traffic.
3) **Atleus:** Atleus, unlike previous configurations, does not incorporate a 3D-mesh. As explained in section IV, we utilize different interconnect strategies for ReRAM and systolic array tiers. Specifically, we utilize an SFC to connect cores in the ReRAM tier given the unidirectional flow of activations among consecutive neural layers. Conversely, we employ a mesh interconnect for the systolic array tier, where both input data and output activations flow. The vertical integration is similar to the second configuration where we introduce additional TSV links (skip links) connecting the top and bottom tier to support intra-layer traffic.

Fig. 8(a) presents the number of routers with a specific number of ports, for the above-mentioned NoC topologies. The router port distribution for the second configuration leads to a higher number of bigger router ports given the integration of additional links to a mesh NoC. In contrast, we observe a lateral shift to lower router port count in Atleus when compared to 3D-mesh. This shift can be attributed to the reduced number of horizontal links by connecting ReRAM cores in an SFC within the ReRAM tiers. Specifically, each core has a planar link only to its immediate predecessor and successor along the SFC path.

We now determine the cost of utilizing the above mentioned NoC configurations. The cost of a single die (denoted as $C_{die}$) can be estimates as:

$$C_{die} = \left(\frac{C_{wafer}}{N_{die}}\right) / Y_{die} \quad (6)$$

where $C_{wafer}$ is the cost of the wafer, $Y_{die}$ is the die yield representing the fraction of working dies from the total number of dies on a wafer, and $N_{die}$ is the number of dies per wafer [50]. $N_{die}$ is given as:

$$N_{die} = \frac{\pi \times \left(\phi_{wafer}/2\right)^2}{A_{die}} - \frac{\pi \times \phi_{wafer}}{\sqrt{2 \times A_{die}}} \quad (7)$$

Here $\phi_{wafer}$ is the wafer diameter and $A_{die}$ is the area of die under consideration. The die yield, $Y_{die}$, depends on the defect density ($D_0$) and the defect clustering ratio ($\alpha$) across the wafer [50]. It is calculated as:

$$Y_{die} = Y_{wafer} \times \left(1 + \frac{A_{die} D_0}{\alpha}\right)^{-\alpha} \quad (8)$$

Assuming the yield for all the dies on the wafer is the same, the cost of a 3D-IC ($C_{3D}$) can be determined by summing the costs of individual dies and accounting for the yield of the stacking process ($Y_{stacking}$) and the TSV yield ($Y_{TSV}$) [51]. The equation for the cost of a 3D-IC is:

$$C_{3D} = \frac{\sum_{i=1}^{n} C_i}{Y_{stacking}^{n-1} Y_{TSV}} \quad (9)$$

where $n$ is the number of stacked dies [51].

We utilize a normalized cost metric to compare the cost of each die. Since the three systems considered with the above-mentioned NoC configurations utilize the same number of stacks, the normalized cost of die ($die\_A$) with respect to ($die\_B$) can be defined as:

$$C_{norm}(die\_A|die\_B) = \frac{Y_{die-B} \times N_{die-B}}{Y_{die-A} \times N_{die-A}} \quad (10)$$

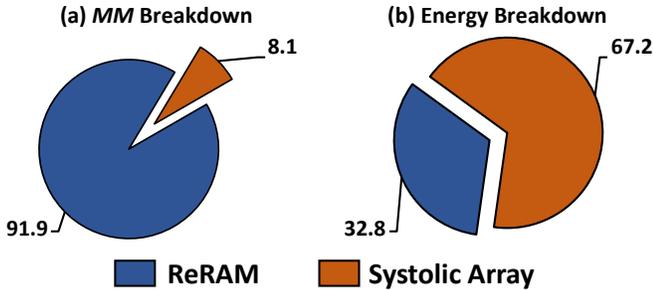

Figure 7: The compute operations and energy breakdown among ReRAM and systolic array in percentage within Atleus for GPT-2 (Medium).

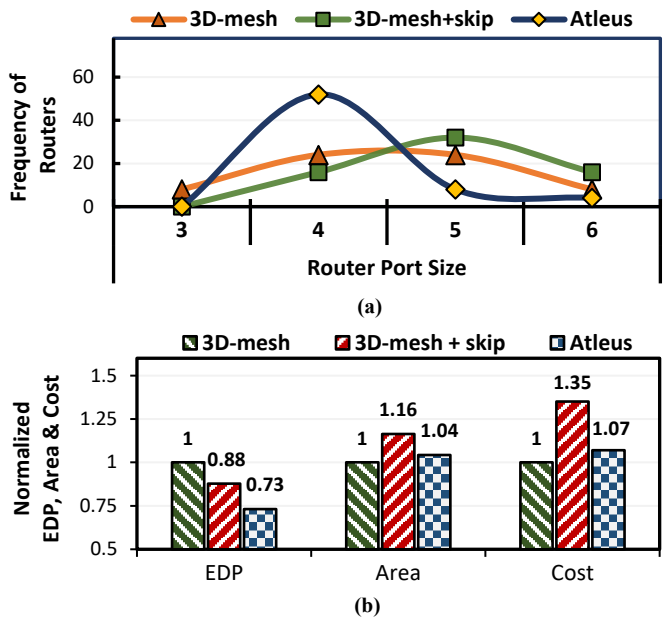

**Figure 8: Comparative evaluation assessing (a) router port size, (b) Normalized EDP, area & cost compared to the baseline 3D-mesh NoC.**

The normalized cost for our analysis becomes a function of $D_0$, $\alpha$, $\phi_{wafer}$, and $A_{die}$. Here, $D_0$, $\alpha$, and $\phi_{wafer}$ are physical parameters. The only controllable parameter influencing the cost is the area of each die.

The die area ($A_{die}$) is made up of three components :

$$A_{die} = A_{core} + A_{router} + X_{TSV}A_{TSV} \quad (11)$$

where $A_{core}$ is the area of each core, $A_{router}$ is the area occupied by NoC routers, and $X_{TSV}$ is the number of TSVs, and $A_{TSV}$ is the area per TSV considering the TSV pitch. The pitch is defined the distance between the center of two adjacent TSVs [52]. Note that as TSVs block routable area, $X_{TSV}A_{TSV}$ is not available for metal interconnect [50] [53].

For our analysis, we assume $pitch = 3 \times Diameter_{tsv}$ [52], $D_0 = 0.2$ [53], $\alpha = 3$ [53], $\phi_{wafer} = 300nm$ [50] [51] [53]. Each NoC design influences the area of router ($A_{router}$) and the area occupied by TSVs ($X_{TSV}A_{TSV}$) with core area ($A_{core}$) remaining consistent. Utilizing Eq. 10, we determine the normalized cost of a die utilizing the different NoC configurations considered with respect to the 3D-mesh system. Fig. 8(b) shows the energy-delay product (EDP), the associated area, and the cost for the 3D system with above-mentioned NoC configurations. We employ EDP as the NoC performance evaluation metric since it captures both performance and energy in a single term. From Fig. 8(b), we observe that integrating additional TSV links to a mesh NoC leads to EDP improvement by 12% but also results in an increase in area by 16% and cost by 35.5%. In contrast, the NoC in Atleus significantly outperforms the baseline mesh with a minimal increase in area. Specifically, we observe a 27% reduction in EDP with only a

4% increase in the NoC area and a 7.5% increase in cost. The use of additional skip links helps in energy-efficient skip-tier communication. Moreover, the reduced number of horizontal links helps to offset the area overhead incurred in integrating additional skip links. This makes the 3D NoC in Atleus an equally practical alternative relative to the conventional and widely used 3D-mesh enabled architecture without imposing a prohibitive area overhead or excessive cost.

**Cost comparison with a 2D architecture:** Atleus employs a four planar-tier 3D manycore architecture, with 16 cores placed within each $100 \; mm^2$ tier, as illustrated in Fig. 2. Existing work has shown that the dominating factor in determining the overall cost of a system is the cost of the dies, which is influenced by the high wafer costs [50] [51] [52] [53]. The cost of a die is a function of the die yield, which depends on the area of the die, as shown in Eqs. 6, 7 and 8 above. We observe that the combined die cost of the four tiers in Atleus, each of $100 \; mm^2$, is lower than the cost of the 2D baseline die of equivalent area ($400 \; mm^2$). Specifically, the die cost of the 2D baseline is approximately 67% higher than that of Atleus. This cost advantage is primarily due to the improved die yield enabled by the smaller area of each tier in Atleus. The enhanced die yield helps reducing the additional cost overheads associated with 3D integration, including the yield challenges of through-silicon vias (TSVs) and vertical stacking processes ($Y_{TSV}$ and $Y_{stacking}$). While this comparison focuses on only die cost which is the dominating cost in a system, a comprehensive end-to-end cost comparison between the 2D and 3D systems would include additional factors such as assembly costs, thermal cooling costs, off-chip HBM costs, testing costs, additional I/O costs, and packaging costs [50] [51]. However, these considerations fall outside the scope of this paper.

### E. Noise-Aware Finetuning

Atleus employs ReRAM cores for the pre-trained weights while utilizing systolic array cores for the trainable LoRA matrices. As discussed in Section 2, ReRAM crossbars are susceptible to noise, which can negatively affect model accuracy. Therefore, it is crucial to address the impact of ReRAM non-idealities to ensure reliable fine-tuning within the Atleus framework. For this purpose, we follow existing work and utilize a noise injection approach to improve the robustness of the finetuned model to ReRAM non-idealities [54] [55]. It is important to highlight that any other noise-aware training algorithm can also be applied for this purpose [14] [56]. As previously mentioned, the training of LoRA weights in Atleus (exclusively) utilizes solely systolic array cores during fine-tuning. Consequently, unlike the pre-trained weights stored on ReRAM crossbars, the LoRA weights do not get affected by the non-idealities associated with ReRAM devices. The output activations in Atleus are the sum of activations from pre-trained weights and the LoRA weights. Therefore, we inject Gaussian noise into the pre-trained weights while training the LoRA weights in the presence of this noise during the fine-tuning process, making the overall model resilient to noise. This approach ensures that the trained LoRA weights for each downstream task are robust to the noise in the pre-trained weights stored on ReRAM crossbars.

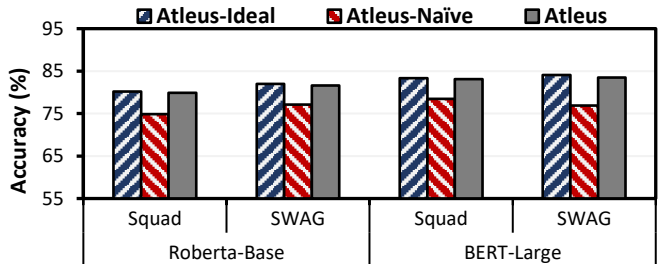

**Figure 9: Comparative analysis of Atleus with and without noise-aware finetuning against the ideal baseline that assumes no noise.**

We compare the accuracy achieved with and without noise-aware finetuning. For weight variation ($\Delta w$), we adopt a Gaussian distribution with $\Delta w \sim \mathcal{N}(0, \sigma^2)$, where $\sigma$ represents the standard deviation of weights, consistent with prior work [54] [55]. The value of $\sigma$ is determined considering the absolute maximum value of the pre-trained weights in the transformer models used in our evaluation [55]. Any noise-induced weight variations exceeding these bounds are clipped to prevent accuracy degradation [57]. Fig. 9 presents the accuracy of Roberta-Base and Bert-Large models as representative examples of two finetuning tasks. Here, Atleus-Ideal denotes model accuracy assuming ideal ReRAM device behavior and serves as the baseline. Atleus-naïve represents the model accuracy without noise-aware finetuning, while Atleus is the noise-aware finetuning approach. From Fig. 9, we can observe that fine-tuning in the presence of noise effectively restores model accuracy, with negligible accuracy loss (less than 0.5%).

### F. Performance Analysis

We first conduct a comparative analysis between Atleus and another heterogeneous accelerator, HAIMA, during the fine-tuning of transformer models. HAIMA for transformer acceleration utilizes a system architecture incorporating a host, SRAM, and DRAM connected via a 2.5D interposer. Within this setup, the host is responsible for executing the score computation, specifically the softmax operation on the output resulting from $Q.K^T$ multiplication executed on HBM [26]. As previously mentioned, Atleus employs a pipelined approach for transformer finetuning. The intra-layer pipeline consists of four stages, denoted as $S_n$, where $n$ represents the stage number. ReRAM cores are employed in three of these stages (i.e., $S_1$, $S_3$ and $S_4$), while systolic array is utilized in the remaining stage ($S_2$). Fig. 10 illustrates the computation and communication latency for each pipeline stage during finetuning of the BERT-Large model with the SWAG dataset as a representative example. We account for both communication and computation latencies for each pipeline stage since compute speed-up needs to be complemented with low-latency interconnect for end-to-end performance gain. From Fig. 10, we observe a significantly high communication delay during the MHA and LoRA calculation for HAIMA. This can be attributed to multiple transformer layers sharing the same interposer links for data exchange with the host or SRAM cores leading to congestion. For instance, a single host device is used for score calculation, resulting in many-to-one traffic. We see a similar scenario for $K$, $Q$ and $V$ computation since HBM is utilized for determining $K$ and $Q$ values, whereas $V$ is computed on SRAM cores [26]. In contrast, Atleus utilizes an integrated 3D architecture using vertical links to enable efficient communication among the heterogeneous cores.

Further, we see higher computation time for HAIMA in comparison to Atleus. As mentioned earlier, the pipelined architecture executes transformer encoder/decoder layers concurrently across different stages. However, the HBM design limits the number of parallel bank accesses for power management purpose [58]. This limits the parallel execution of layers when utilizing the HBM. Now, the computation on HBM must be multiplexed during the fine-tuning process which leads to higher computation delay. This bottleneck is particularly evident in Fig. 10 during the FF computation which involves multiplication with large weight matrices on HBM. Conversely, Alteus effectively accelerates the FF computation with parallel execution on ReRAM cores. From Fig. 10, we see that the stage delay in HAIMA is significantly influenced by both communication delay across heterogeneous computing cores and the parallel computation across pipeline stages. This makes HAIMA unsuitable for pipelining and transformer fine-tuning. Overall, Atleus achieves a notably lower stage delay in comparison to HAIMA, leading to overall speedup.

Next, we compare the end-to-end latency of Atleus with respect to the baselines HAIMA, 3D-TPU, and GPU. Figs. 11(a) and 11(b) show the normalized execution time and energy respectively for Roberta-Base and BERT-Large models as examples on two fine-tuning tasks/datasets. We utilize the stage delay determined in Fig. 10 for reporting end-to-end latency in HAIMA and Atleus. We observe noticeable limitations in HAIMA, 3D-TPU, and GPU for transformer fine-tuning. HAIMA suffers from high latency due to higher stage delay as shown in Fig. 10. On the other hand, GPU systems process instructions serially, one after another. The parallelism within GPU is confined to single computing kernels distributed across multiple GPU cores. Further, we observe less than 50% GPU core utilization highlighting the memory bottleneck. 3D-TPU performs ~2× faster in comparison to GPUs as the 3D architecture performs better with the complex dataflow within transformers during fine-tuning. Additionally, a more effective pipelined implementation supports better performance in 3D-TPU. However, 3D-TPU also suffers from low average utilization and high data transfer costs from memory. Conversely, Atleus outperforms GPU, 3D-TPU, and HAIMA, achieving consistent speedup through an efficient pipelined design that ensures high throughput. Further, ReRAM cores perform MVM operations in an energy-efficient manner without the need to fetch data from off-chip.

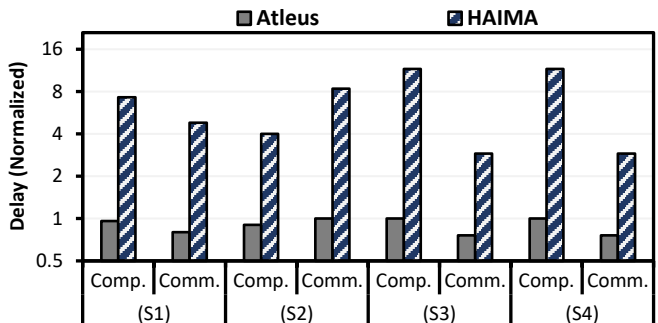

**Figure 10: Pipeline stage delay for HAIMA and Atleus for finetuning BERT-Large model with SWAG dataset. The labels S1, S2, S3, and S4 denote the four respective pipeline stages.**

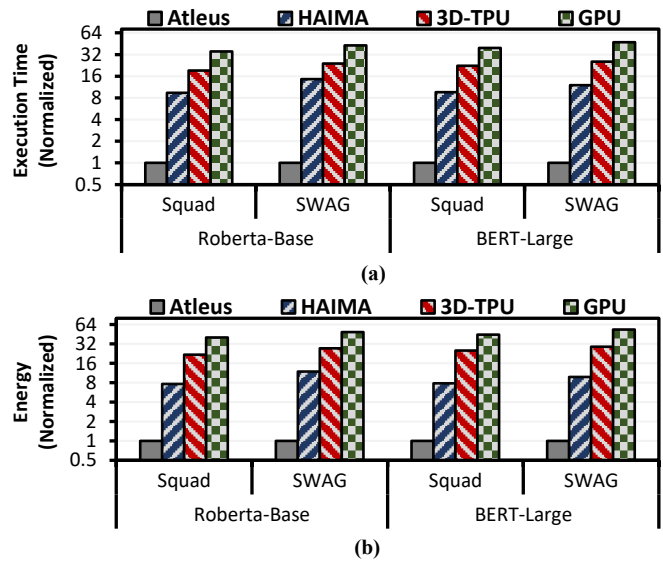

**Fig. 11: Normalized execution time and energy when compared to the baseline Atleus architecture for HAIMA and GPU.**

### F. Quantization Evaluation

Finally, we evaluate the effectiveness of crossbar-wise quantization on Atleus. We quantize the pre-trained model via crossbar-wise quantization with different bit precision and determine the impact during fine-tuning. We use two transformer models, GPT-2 (Medium) and BLOOM-560m for the evaluation and fine-tuning on the language modeling dataset "Wikitext". Each quantization configuration is denoted as M$n$F$m$, where "$n$" and "$m$" represent the number of bits utilized for the MHA and FF module, respectively. Note that the LoRA parameters are not quantized during fine-tuning. Here, we utilize a 16-bit representation as the baseline for model weights. We first compare the energy for Atleus against the baselines for the different quantization configurations. Fig. 12 illustrates the energy trendline for each architecture, normalized to their respective 16-bit implementations for BLOOM-560m as an example with different quantization bits. The trend is similar across other datasets. We observe almost a linear decrease in energy consumption for Atleus characterized by a slope, $\alpha$, where $(\alpha < 1)$. The value of $\alpha$ is less than 1 since dequantization adds overhead to the ReRAM tile architecture. Furthermore, there is no observable reduction in NoC power since activations are dequantized at the tile level. Notably, the energy savings are higher for M8F4 in comparison to M4F8 since a lower precision for the FF network results in more resource saving compared to MHA. Conversely, we see that GPU energy increases upon quantization in Fig. 12. This increase is attributed to the increase in execution time when using a quantized pre-trained model. The model is dequantized once fetched from memory to align with GPU compute precision. We see similar observations for 3D-TPU where overall system energy increases after quantization. HAIMA also observes an increase in energy with the dequantization operation necessary for the model to be executed within DRAM banks. Note that DRAM is a hard IP and integration of additional circuitry for dequantization is not straightforward.

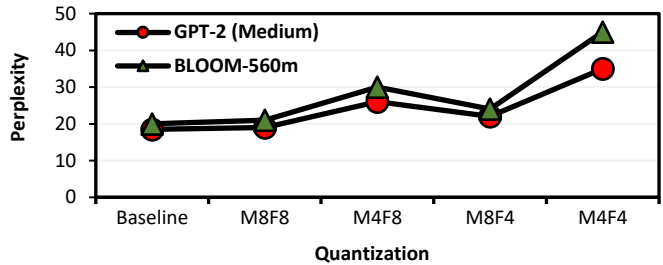

Figure 13: Model perplexity with different quantization configurations for GPT-2 (Medium) and BLOOM-560m on Atleus.

Next, we quantify the transformer model performance under different quantization configurations. We use perplexity as the evaluation metric. It is a measure of the model's predictive accuracy on a given dataset and is derived by taking the exponential of the model's loss. Fig. 13 illustrates the model perplexity with different quantization bits used for pre-trained models. A lower perplexity indicates a better performing model. From Fig. 13, we observe that M8F8 does not result in any notable increase in perplexity. The M8F4 quantization exhibits a slight increase in perplexity, though still indicating stable model performance. In contrast, the perplexity increase is noticeable for the M4F8. This is attributed to the greater dynamic range observed in the weights in MHA when compared to FF weights during quantization. Overall, we observe that M8F4 outperforms M4F8 in terms of both model stability and energy efficiency. The fine-tuning fails with the M4F4, where we observe a significant increase in model perplexity. This is primarily attributed to a single quantization constant used for a larger count of weights. Each ReRAM crossbar is of the size 128*128, with a 2-bit representation per cell. When a 4-bit representation is utilized in M4F4, approximately 8,000 numerical values are quantized using a single constant. This results in suboptimal representation where certain bins are underutilized or completely unused. To summarize, both M8F8 and M8F4 ensure model stability and one of these configurations can be chosen based on the desired performance and energy trade-off as shown in Fig. 12 and Fig. 13. The M8F8 is preferable when the model performance is of primary concern since it maintains model performance (Fig. 13). Conversely, the M8F4 is suitable when energy efficiency is of paramount importance, accepting a slight compromise in model performance (approximately 2-3%). Finally, Fig.14 shows the normalized energy in comparison to Atleus showing increased gains upon quantization for transformer models GPT-2 (Medium) and BLOOM-560m. It is evident that Atleus' energy reduces as the quantization level increases. In contrast, HAIMA, 3D-TPU, and GPU demonstrate an increase in energy consumption under similar conditions. This differential behavior underscores the efficiency of the Atleus architecture enabled by crossbar-wise quantization.

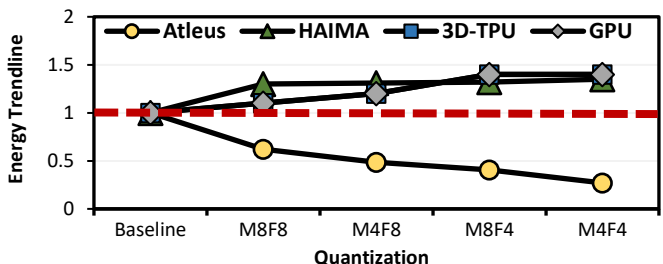

Figure 12: Comparative Analysis of system energy for each hardware architecture under different quantization configuration normalized to their respective 16-bit implementation.

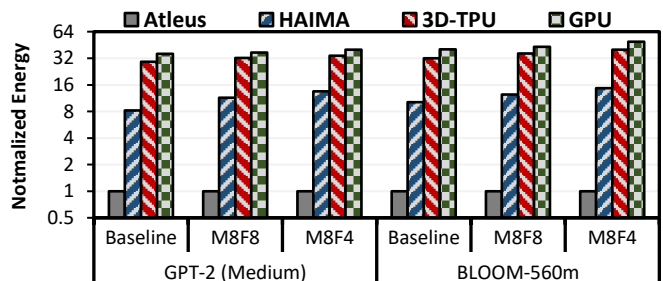

Figure 14: Normalized Energy as compared to the baseline Atleus for different quantization bit precisions.

### G. Inferencing on Atleus

Finally, we analyze the effectiveness of Atleus for transformer inference to evaluate its general utility across both edge deployment scenarios. We utilize the fine-tuned models trained with LoRA for this analysis. The computational kernel mapping remains the same as fine-tuning where the pre-trained model weights are stored on ReRAM cores and the trained LoRA parameters are stored off-chip in DRAM. The only difference is that there is no weight update for the LoRA parameters during inference. Fig. 15 shows the normalized execution time for Roberta-Base and BERT-Large on the evaluation dataset of Squad and SWAG after fine-tuning as examples. We see a similar trend as that of transformer fine-tuning with Atleus outperforming HAIMA, 3D-TPU, and GPU. A key advantage of storing the pre-trained transformer model on ReRAM cores is the energy efficiency when using the same transformer model fine-tuned on different tasks. Now, inferencing can be executed on different tasks by just loading LoRA parameters from off-chip, which constitutes only a fraction of the pre-trained model parameters. This is not possible for both GPU and 3D-TPU as they load the complete model along with LoRA parameters. HAIMA performs slightly better compared to fine-tuning given the simplified traffic during inference. However, it still utilizes the host and the SRAM module for computations in a 2.5D system. Overall, the findings confirm that Atleus also outperforms existing accelerators on transformer inference tasks.

## VI. Conclusion

Transformer models, with their intricate self-attention mechanisms and substantial memory demands, pose significant challenges for acceleration on conventional computing architectures. In this paper, we design a domain-specific 3D manycore heterogeneous architecture called Atleus for deploying transformers on the edge. Atleus, unlike existing transformer accelerators, works for both transformer fine-tuning and inference. Atleus implements an intra-layer pipeline across heterogeneous NVM and systolic array computing cores to enhance overall system throughput. Importantly, Atleus leverages the advantage of NVM devices while overcoming the reliability challenges associated with frequent rewrites. Further, Atleus employs a NoC to support the dataflow in transformers and achieve high performance. Finally, Atleus enables crossbar-wise quantization to further enhance energy efficiency. Experimental results demonstrate that Atleus outperforms existing state-of-the-art by up to 56× in speedup and 64.5× in energy efficiency compared to GPU.

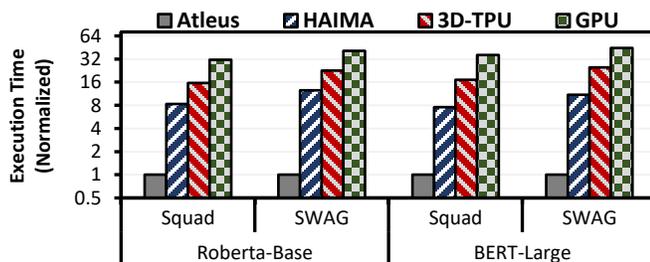

**Figure 15: Normalized inference execution time when compared to the baseline Atleus for different transformer models.**

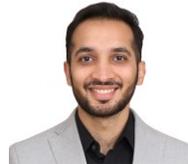
**Pratyush Dhingra** (Graduate Student Member, IEEE) received the B.E degree in electronics and computer engineering from NSIT, Delhi University, Delhi, India, in 2019. He is currently pursuing the Ph.D. degree in Computer Engineering at Washington State University, Pullman, WA, USA.

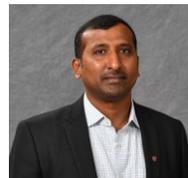
**Janardhan Rao Doppa** (Senior Member, IEEE) received the Ph.D. degree in computer science from Oregon State University, Corvallis, OR, USA, in 2014. He is the Huie-Rogers Endowed Chair Associate Professor in Computer Science at Washington State University, Pullman, WA, USA.

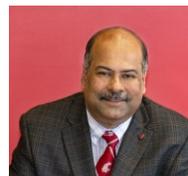
**Partha Pratim Pande** (Fellow, IEEE) received the Ph.D. degree in electrical and computer engineering from the University of British Columbia, Vancouver, BC, Canada, in 2005. He is a professor and a holder of the Boeing Centennial Chair of Computer Engineering with the School of Electrical Engineering and Computer Science, Washington State University, Pullman, WA, USA.